\documentclass[useAMS,usenatbib]{mn2e} 

\usepackage{times}
\usepackage{graphicx}
\usepackage{amsmath}
\usepackage{float}
\usepackage{ifthen}
\usepackage{times}
\usepackage{natbib}
\usepackage{rotating}
\usepackage{units}

\def\gsim{\mathrel{\raise0.35ex\hbox{$\scriptstyle >$}\kern-0.6em
\lower0.40ex\hbox{{$\scriptstyle \sim$}}}}
\def\lsim{\mathrel{\raise0.35ex\hbox{$\scriptstyle <$}\kern-0.6em
\lower0.40ex\hbox{{$\scriptstyle \sim$}}}}

\begin{document}

\title[Clustering of HAEs at $z=2.23$]{The clustering of H$\alpha$ emitters at
$\mathbf{z=2.23}$ from HiZELS} \author[J. E. Geach et
al.]{\parbox[h]{\textwidth}{ J.\ E.\ Geach,$^1$\thanks{E-mail:
jimgeach@physics.mcgill.ca} D.\ Sobral$^2$, R.\ C.\ Hickox$^3$, D.\ A.\
Wake$^4$, Ian\ Smail$^5$, P.\ N.\ Best$^6$, C.\ M.\ Baugh$^5$ and J.\ P.\
Stott$^5$} \vspace*{6pt}\\\noindent $^1$Department of Physics, McGill
University, 3600 Rue University, Montr\'eal, Qu\'ebec, H3A 2T8,
Canada\\$^2$Leiden Observatory, Leiden University, P.O. Box 9513, NL--2300 RA
Leiden, The Netherlands\\$^3$Department of Physics and Astronomy, Dartmouth
College, 6127 Wilder Laboratory, Hanover, NH 03755, USA\\$^4$Department of
Astronomy, Yale University, New Haven, CT 06520, USA\\$^5$Institute for
Computational Cosmology, Department of Physics, Durham University, South Road,
Durham, DH1 3LE, UK\\$^6$SUPA, Institute for Astronomy, Royal Observatory of
Edinburgh, Blackford Hill, Edinburgh, EH9 3HJ, UK\\}

\date{}

\pagerange{\pageref{firstpage}--\pageref{lastpage}} \pubyear{2012}

\maketitle

\label{firstpage}

\begin{abstract}We present a clustering analysis of 370 high-confidence
H$\alpha$ emitters (HAEs) at $z=2.23$. The HAEs are detected in the Hi-Z
Emission Line Survey (HiZELS), a large-area blank field 2.121$\mu$m narrowband
survey using the United Kingdom Infrared Telescope (UKIRT) Wide Field Camera
(WFCAM). Averaging the two-point correlation function of HAEs in two
$\sim$1\,degree scale fields (United Kingdom Infrared Deep Sky Survey/Ultra
Deep Survey [UDS] and Cosmological Evolution Survey [COSMOS] fields) we find a
clustering amplitude equivalent to a correlation length of
$r_0=3.7\pm0.3$\,$h^{-1}$\,Mpc for galaxies with star formation rates of
$\gsim$7\,$M_\odot$\,yr$^{-1}$. The data are well-fitted by the expected
correlation function of Cold Dark Matter, scaled by a bias factor:
$\omega_{\rm HAE}=b^2\omega_{\rm DM}$ where $b=2.4^{+0.1}_{-0.2}$. The
corresponding `characteristic' mass for the halos hosting HAEs is $\log
(M_{\rm h}/[h^{-1}M_\odot])=11.7\pm0.1$. Comparing to the latest semi-analytic
{\sc galform} predictions for the evolution of HAEs in a $\Lambda$CDM
cosmology, we find broad agreement with the observations, with {\sc galform}
predicting a HAE correlation length of $\sim$4\,$h^{-1}$\,Mpc. Motivated by
this agreement, we exploit the simulations to construct a parametric model of
the halo occupation distribution (HOD) of HAEs, and use this to fit the
observed clustering. Our best-fitting HOD can adequately reproduce the
observed angular clustering of HAEs, yielding an effective halo mass and bias
in agreement with that derived from the scaled $\omega_{\rm DM}$ fit, but with
the relatively small sample size the current data provide a poor constraint on
the HOD. However, we argue that this approach provides interesting hints into
the nature of the relationship between star-forming galaxies and the matter
field, including insights into the efficiency of star formation in massive
halos. Our results support the broad picture that `typical' ($\lsim$$L^\star$)
star-forming galaxies have been hosted by dark matter haloes with $M_{\rm
h}\lsim10^{12}h^{-1}M_\odot$ since $z\approx2$, but with a broad occupation
distribution and clustering that is likely to be a strong function of
luminosity.

\end{abstract} 

\section{Introduction}

The Cold Dark Matter model contends that galaxies are biased tracers of an
unseen, underlying cold dark matter distribution that has evolved from
primordial fluctuations into a rich hierarchy of structure, with baryons
forming into galaxies within gravitationally bound dark matter halos (White \&
Rees\ 1978). Understanding the relationship between the distribution of
observed galaxies, their properties, and their co-evolution with the latent
matter field is a key question of observational cosmology, and can yield
important information about a galaxy population (Peebles\ 1980).

One of the simplest, but also the most powerful, tools at our disposal to
address this issue is the clustering of galaxies, as has been recognised for
many years (Rubin\ 1954; Groth \& Peebles\ 1977; Peebles\ 1980). At a basic
level, the statistics of counts of galaxy pairs, relative to random
distributions, reveal the scales over which the fluctuations in the spatial
distribution of galaxies are correlated, and therefore a measure of how
`clustered' a population is; longer correlation lengths correspond to stronger
clustering and an indication that those galaxies are hosted by, on average,
more biased and hence more massive dark matter halos (e.g.\ Mo \& White\ 1996).

In the local Universe, mature wide-area surveys such as the Sloan Digital Sky
Survey (SDSS; York et al. 2000) and the Two Degree Field (2dF) Redshift Survey
(Colless et al.\ 2001), have delivered highly accurate measurements of the
clustering of populations of galaxies and quasars (Norberg et al. 2001; Myers
et al.\ 2006; Ross et al.\ 2009; Wake et al.\ 2008; Zehavi et al.\ 2011). A
key result of these studies is the observation that the clustering amplitude
is enhanced as the mass limit of the galaxy sample increases, indicating that
the more massive galaxies are hosted by more massive halos. Furthermore, it is
clear that {\it passive} galaxies are more strongly clustered on small spatial
scales compared to galaxies with ongoing star formation (e.g.\ Norberg et al.\
2002). Over the past decade a method of interpreting these observations has
been developed (in part motivated by large {\it N}-body simulations) which
expresses the distribution of galaxies relative to the matter field through a
probabilistic halo occupation distribution (HOD; Benson et al.\ 2000; Cooray
\& Sheth\ 2002; Zheng et al.\ 2005) or, similarly, a conditional luminosity
function (Yang et al.\ 2003). Halo models provide an intuitive framework to
relate observed projected correlation functions to the hierarchical paradigm,
and are becoming increasingly common tools for the interpretation of
clustering data.

Clustering analyses are now routine for high redshift ($z>1$) mass-limited
galaxy samples, largely thanks to the increased efficiency of deep and
wide-area ($\sim$1\,degree scale) multi-band
(ultraviolet--optical--near-infrared) imaging surveys offering excellent
photometric redshifts (accurate to the few percent level at $z\sim1$) and
stellar mass estimates for large numbers of massive galaxies (e.g.\ Wake et
al.\ 2011). When it comes to measuring the clustering properties of purely
star-forming galaxies at high redshifts, which -- in the halo model context --
could yield important clues about the environmental trends in the history of
stellar mass assembly, the main challenge is to understand the selection
function, since most broad-band selections (Lyman Break, BX/BM, `s{BzK}', and
so-on) can result in heterogeneous samples with broad redshift distributions,
and can be biased towards stellar mass in complicated ways. The latter two
issues are undesirable, given the strong evolution in the specific star
formation rates (SFRs) of galaxies since $z\sim1$--2 (Noeske et al.\ 2007;
Elbaz et al.\ 2011).

Narrowband ($\Delta \lambda/\lambda\simeq10^{-3}$) selections of star-forming
galaxies are of great value in this regard, as they allow for the clean
selection of galaxies based simply on the strength of an emission line sampled
by the filter. The narrow bandpass corresponds to a narrow redshift window,
within which the population is not expected to evolve. The main contaminants
to such a survey are emission-line galaxies at different redshifts
corresponding to the redshifting of alternative lines into the band. For
high-{\it z} surveys these contaminants are predominantly lower-redshift
populations and easily removed (see \S2). Most narrowband-selected clustering
analyses conducted so-far have targeted the Ly$\alpha$ emission line,
redshifted into the optical window for $z\simeq3$ and thus convenient for
deep, wide-field surveys out to very high redshifts (e.g.\ Ouchi et al.\
2003). The development of wide-format infrared cameras over the past decade
has now cleared the way for panoramic near-infrared narrowband surveys that
target the H$\alpha$ nebular line at epochs of $z\sim$1--2, spanning the peak
in the global star formation rate density, and thus one of the most important
intervals in galaxy formation studies. H$\alpha$ is favoured over the
Ly$\alpha$ line because of its (a) weaker dust obscuration (and ease of
extinction correction, if the Balmer decrement is known), (b) better
understood radiative transfer compared to the resonant Ly$\alpha$ and (c) more
accurate luminosity-to-star formation rate calibrations from surveys of local
star forming regions. It is also important to measure the clustering of HAEs
in preparation for the {\it Euclid} mission, as one of the probes used to
constrain the nature of dark energy will be a slitless redshift survey of HAEs
(Laureijs et al.\ 2011).

In this paper we present a clustering analysis of H$\alpha$ emitters (HAEs) at
$z=2.23$ detected in our Hi-Z Emission Line (HiZELS) survey: a wide-field
near-infrared narrowband survey selecting H$\alpha$ emitting galaxies in three
narrow `slices' of redshift at $z=0.84$, $z=1.47$ and $z=2.23$ (e.g.\ Geach et
al\ 2008; Best et al.\ 2010; Sobral et al.\ 2010, 2012). In \S2 we provide a
brief review of the observations and selection technique (although we refer
the reader to the aforementioned HiZELS publications for a complete,
comprehensive description); in \S3 we describe the clustering analysis and
present the results in \S4, where we approach the interpretation of the data
with a series of models of increasing sophistication, from a simple power law
fit to a full halo model. In \S5 we discuss our findings and conclude with a
review of the main results in \S6. Throughout this work we quote magnitudes on
the AB system, and assume a cosmology with $\Omega_{\rm m}=0.27$, $\Omega_{\rm
\Lambda}=0.73$, $\sigma_8=0.8$ and $H_0=100h$\,km\,s$^{-1}$\,Mpc$^{-1}$ with
$h=0.7$. The co-moving distance to $z=2.23$ is 5128\,Mpc in this cosmology.

\section{Narrowband selection of H$\alpha$ emitters}

The observations and selection of HAEs in the primary HiZELS fields of the
United Kingdom Infrared Deep Sky Survey (UKIDSS) Ultra Deep Survey (UDS;
Lawrence et al.\ 2007) and Cosmological Evolution Survey (COSMOS; Scoville et
al.\ 2007) are described in more detail by Sobral et al.\ (2012) -- we refer
the reader to that article for a comprehensive overview of the selection
technique, but in short we first select galaxies based on the significance of
their `colour excess' in the narrow band. Corrections to the continuum slope
over the bandpass of the {\it K}-band filter (which could mimic a colour
excess) is performed by interpolating over the neighbouring broad band (in
this case, the $H$-band). Further broad band colour selections are performed
to refine the selection (which can be contaminated by lower redshift Paschen
and Brackett lines for example). Here we perform a flux cut to obtain a
catalogue of approximately uniform depth across both UDS and COSMOS fields.

The flux limit at which we are uniformly complete to $>$50\% over both UDS and
COSMOS fields is $f_{\rm H\alpha} =
5\times10^{-17}$\,erg\,s$^{-1}$\,cm$^{-2}$, corresponding to a luminosity of
$\log_{10}(L_{\rm H\alpha}/{\rm erg\,s^{-1}})=42.3$ at $z=2.23$. Note that
variations in the exact depth of each WFCAM pointing (each field is a mosaic
of several pointings) corresponds to a variation in the surface density of
galaxies. The impact of this on our measured clustering is in part absorbed
into the error bars calculated by jackknife resampling of the survey area that
we describe in \S3.2. Assuming a $L_{\rm H\alpha}$--SFR calibration of
$1.3\times10^{41}$\,erg\,s$^{-1}$ per $M_\odot$\,yr$^{-1}$ (Kennicutt et al.\
1998), our selection is SFR limited at $\geq$7\,$M_\odot$\,yr$^{-1}$ assuming
a canonical 1\,mag of extinction in the H$\alpha$ line. Foreground sources are
easily removed by high-quality photometric redshifts estimated from
UV--optical--near-infrared photometry in both the UDS and COSMOS fields.
Sobral et al.\ (2012) present the $z_{\rm phot}$ distribution for $K$-band
selected HAEs, indicating the most significant peak in the distribution at
$z=2.23$, but with low-redshift enhancements at the expected wavelengths of
Pa$\alpha$, Pa$\beta$, He\,{\sc i}, [S\,{\sc iii}], and at high redshift
[O\,{\sc iii}] at $z\sim3.3$.

To refine the photometric selection, we make use of a key design feature of
HiZELS, namely the fact that our custom-made {\it J} and {\it H}-band
narrow-band filters select {[O\,{\sc ii}]} and {[O\,{\sc iii}]} emitters at
$z=2.23$ respectively. Thus, double or triple detections for the same source
in each of the narrow-bands provides an extremely robust selection with almost
no contamination. There are 84 $z=2.23$ HiZELS sources detected in this way,
and this is used to refine photometric redshift cuts and broad-band
photometric selections as described in further detail by Sobral et al.\
(2012). In summary, the overall contamination rate from non-HAEs in our sample is expected to be $\lsim$10\%.

The total number of galaxies detected in each field satisfying these selection
criteria is 230 and 140 HAEs in COSMOS and UDS respectively. The higher number
of HAEs in the COSMOS field is due to the difference in survey areas: HiZELS
has so-far covered 1.23\ deg$^2$ in COSMOS and 0.75\,deg$^2$ in UDS. Note that
the surface density of HAEs measured in the two independent fields is nearly
identical, $\Sigma_{\rm HAE}$$\approx$190\,deg$^{-2}$.

\begin{figure*}
\includegraphics[width=0.5\textwidth,angle=-90]{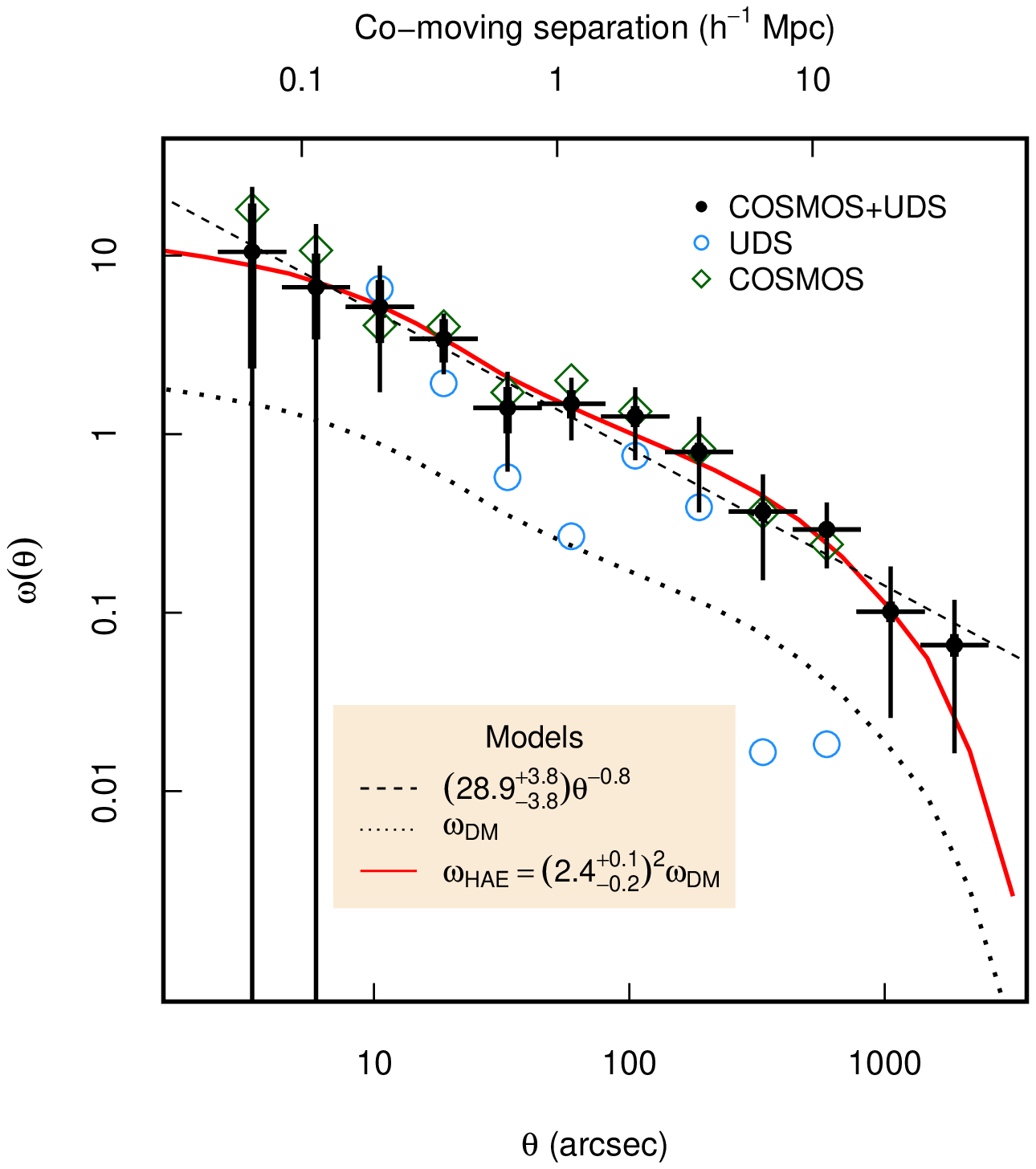}\includegraphics[width=0.5\textwidth,angle=-90]{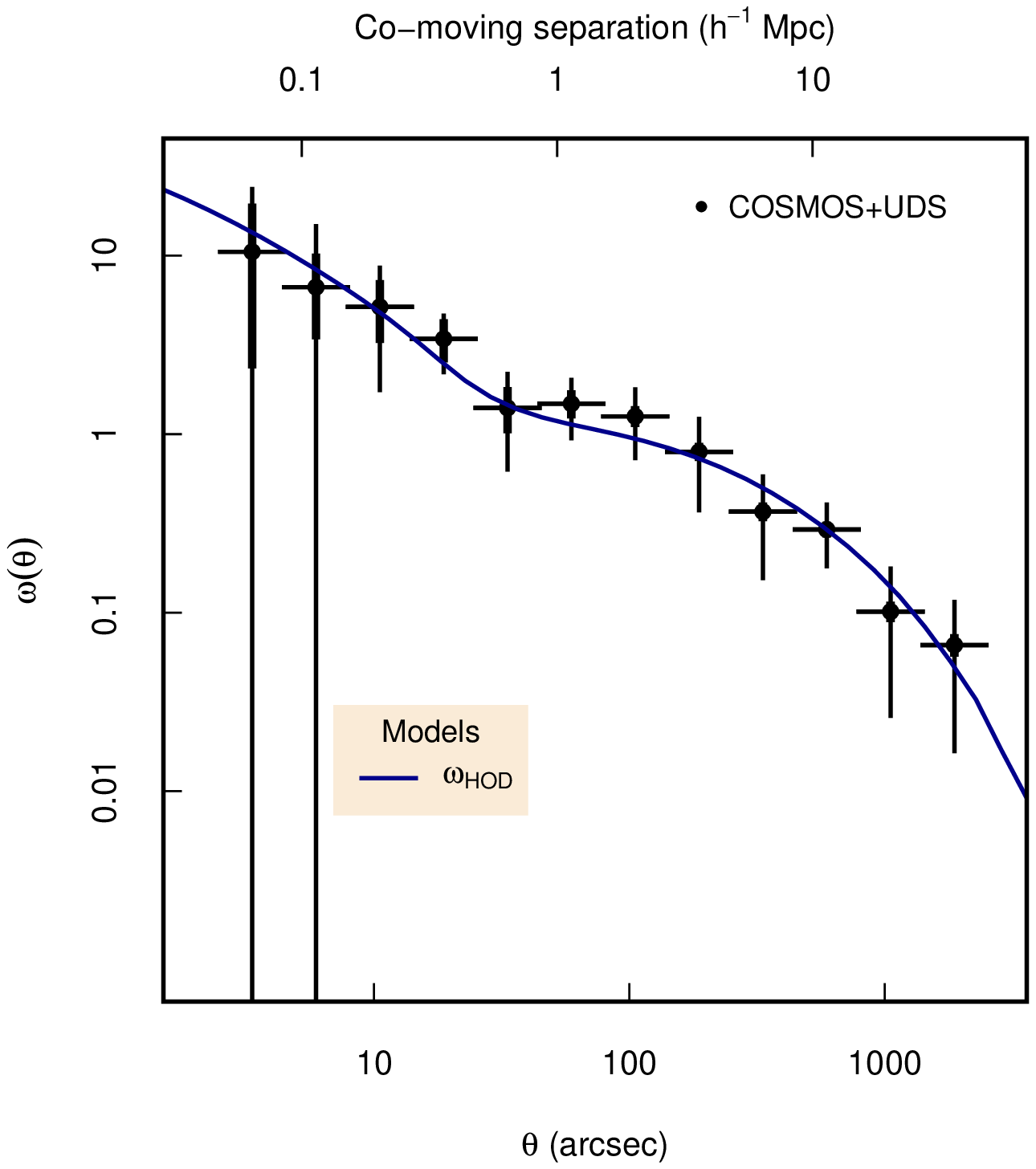}
\caption{Two-point angular correlation function of HAEs in the COSMOS and UDS
fields. (left) We show two model fits to the data: (a) a simple power law
$A\theta^{-0.8}$ (dashed line) and (b) the projected correlation function of
dark matter, scaled by a bias parameter, $b^2$ (dotted and solid lines). The
power law is a reasonable fit to the general shape of the HAE correlation
function, but the dark matter model also provides a good fit, and captures the
deviation from a simple power law at all scales. The error bars are calculated
from the diagonal elements of the covariance matrix which was estimated from
the jackknife re-sampling method (we show for comparison the Poisson errors as
thicker bars). The correlation function for the individual fields is also
shown, however for clarity we do not show the error bars for these. Note that
the combined COSMOS+UDS $\omega(\theta)$ values have been corrected for the
integral constraint (\S3.1, equation\ 3), whereas the individual fields have
not. (right) Combined correlation function as (left), but shown with the best
fitting HOD model (described in \S4.3). The halo model successfully models the
amplitude of the clustering strength on all measured scales, including the
break at $\sim$1\,$h^{-1}$\,Mpc indicating the transition between the
dominance of the one- and two-halo term in the halo model. } \end{figure*}

\section{Clustering analysis}

\subsection{Two-point angular correlation function estimator}

We calculate the two-point angular correlation function, $\omega(\theta)$,
using the estimator proposed by Landy \& Szalay\ (1993),
\begin{equation} \omega(\theta) = 1 + \left(\frac{N_R}{N_D}\right)^2
\frac{DD(\theta)}{RR(\theta)} - 2\frac{N_R}{N_D}\frac{DR(\theta)}{RR(\theta)},
\end{equation} 

\noindent where $N_D$ and $N_R$ are the number of galaxies in the data and
random catalogue respectively, and $DD$, $RR$ and $DR$ are the number of
data-data, random-random and data-random pairs at angular separation $\theta$.
The modified Poissonian uncertainty is:

\begin{equation}
	\delta\omega(\theta) = \frac{1+\omega(\theta)}{\surd
	DD(\theta)},
\end{equation}

\noindent although this certainly is an underestimate of the true error (we
estimate the full covariance matrix in \S3.2). For the random catalogue, we
distribute $20N_D$ points uniformly over the survey areas, avoiding masked
regions (cross-talk artifacts, bright stellar halos, etc.). We combine the
results from the two independent survey volumes at the pair-counts stage, such
that\ $DD = DD_{\rm UDS}+DD_{\rm COSMOS}$, etc. In practice this gives very
similar results to averaging the individual $w(\theta)$, weighting by the
Poisson uncertainty.

A correction must be applied to $w(\theta)$ due to the finite area surveyed
and the fact that the mean density of galaxies is estimated from the sample
itself and would be biased due to cosmic variance. The integral constraint
($C$; Groth \& Peebles\ 1977) corresponds to a scale-independent
underestimation of $\omega(\theta)$. As in Geach et al.\ (2008), we calculate
$C$ following Roche et al.\ (1999):

\begin{equation}
C = \frac{\sum_i \omega(\theta_i) RR(\theta_i)}{\sum_i
RR(\theta_i)},
\end{equation}

\noindent where we model $\omega(\theta)$ using the scaled angular correlation
function of dark matter, which is an excellent fit to the data and superior to
a single power law (we discuss this analysis in \S4.1). We evaluate equation 3
iteratively: first fitting the model to the data, calculating $C$ and then
applying this correction to the data and fitting again, repeating this process
until there is convergence. We find $C=0.134$ for the combined area, and
correct the measured $\omega(\theta)$ for this factor before fitting models.

\subsection{Error estimation}

We estimate the full covariance using the `delete one jackknife'
method (Shao\ 1986, and see Norberg et al.\ 2009 for a comprehensive
review of this and other error estimation methods). In short, the
survey volume is split into $N$ sub-areas, and $\omega(\theta)$
calculated $N$ times, each time excluding one of the sub-areas.  The
elements of the covariance matrix are then given by: 
\begin{equation}
\mathcal{C}_{ij} = \frac{N-1}{N}\sum^N_{k=1}(\omega^k_i -
\bar{\omega_i})(\omega^k_j - \bar{\omega_j}) '
\end{equation} 

\noindent where $\omega^k_i$ is the correlation function (equation 1) measured
for the $i$th angular bin, for the $k$th jackknife resampling, and

\begin{equation} \bar{\omega_i} = \frac{1}{N}\sum^N_{k=1}\omega^k_i.
\end{equation}

We split the survey volume into 32 sub-regions and evaluate equation 1 for
each jackknife realisation, omitting one sub-region each time. The uncertainty
on the correlation function evaluated at each angular bin is given by
$\delta\omega(\theta_i) = \surd \mathcal{C}_{ii}$ and this is used in the evaluation of $\chi^2$ difference between the data $(\omega)$ and an arbitrary model $(\omega^{\rm
model})$ taking into account covariance is

\begin{equation} \chi^2=(\omega-\omega^{\rm model})^{\rm T}
\mathcal{C}^{-1}(\omega - \omega^{\rm model}), \end{equation}

\noindent with the 1$\sigma$ uncertainty on a model parameter equivalent to
the range $\Delta\chi^2=1$.

\section{Results}

We present the results in Figure\ 1, corrected for the integral constraint,
and including the covariance uncertainties evaluated in equation 4.
Correlation functions are often fitted by a single power law, $\omega(\theta)
= A \theta^{-\beta}$, usually with $\beta\approx0.8$. This is adequate to fit
the overall trend in the data, but the observed correlation function clearly
deviates from a simple power law, especially at $\theta>1'$. In part, the
deviation of the observed correlation function at large separations is due to
the break-down of Limber's approximation at $\theta\gsim600''$ for samples
where $\Delta z$ is narrow (Simon\ 2007, Sobral et al.\ 2010). In this case,
even if the spatial correlation function is a power-law, the angular
correlation function will depart from a power law at large angular
separations. However, we also expect that a single power law is insufficient
to model the clustering across the full angular range for physical reasons
related to the relative clustering of satellite galaxies within single dark
matter halos to the clustering of the halos themselves.

We explore this in the following sections, however for now we start our
analysis with the simple power-law model fitted to data at scales
$\theta\lsim600''$, which is useful for obtaining an estimate of the
correlation length of the galaxies and easily comparable to the clustering of
other populations. We perform minimised $\chi^2$ fits for the amplitude of the
correlation function, fixing the slope with $\beta=0.8$. We find a clustering
amplitude $A = 29\pm4$\,arcsec$^{0.8}$, with a reduced $\chi^2/\nu=0.9$.
Throughout, we quote 1$\sigma$ uncertainties on the $\chi^2$ fit using the
full covariance matrix calculated in equation 6.

If the real space correlation function can be assumed to be $\xi(r) =
(r/r_0)^{-\gamma}$, where $r_0$ is the real-space correlation length and
$\gamma = \beta+1$, the amplitude of the correlation function $A$ can be
related to $r_0$ using a version of Limber's equation (Limber\ 1954; Peebles
1980):

\begin{equation}
A =  r_0^\gamma
\frac{\Gamma((\gamma-1)/2)\Gamma(\gamma/2)}{\Gamma(1/2)}\int_0^\infty 
\frac{H_z}{c}\left(\frac{dn}{dz}\right)^2\chi_z^{1-\gamma}dz,
\end{equation}

\noindent where $A$ is the amplitude of the correlation function evaluated at
$\theta=1$\,radian, $\Gamma$ is the Gamma function, $H_z$ is the Hubble
parameter at redshift $z$, $\chi_z$ is the co-moving radial distance to $z$
and $dn/dz$ is the redshift distribution of the population, normalised to
unity. We assume the redshift distribution of HAEs in our narrowband selection
is set by the H$_2$S(1) filter profile, which can be described by a Gaussian
function centred at $z=2.233$, with full width at half maximum of $\delta
z=0.03$ (e.g.\ Sobral et al.\ 2010). Here we make the further assumption that
we are 100\% incomplete in the wings ($>${\sc fwhm}) of the H$_2$S(1)
transmission function, and therefore define the redshift distribution to be:

\begin{equation}
dn/dz = \left\{ \begin{array}{rl}
n_0\exp(-\frac{(z-z_c)^2}{2\sigma^2} ) & \mbox{for $|z-z_c|<0.015$ } \\
0 & \mbox{for $|z-z_c|\geq0.015$,}
\end{array}\right.
\end{equation}

\noindent where $z_c=2.233$ and $\sigma=0.0126$ and $n_0$ is the normalisation
constant. This form of the redshift distribution attempts to account for the
fact that there is a (luminosity dependent) bias in our selection in favour of
HAEs with observed H$\alpha$ emission closer to the peak transmission of the
filter. We are currently engaged in spectroscopic follow-up projects to
properly characterise the redshift distribution of HAEs in our sample.
Adopting this $dn/dz$ in equation 6, we find $r_0 =3.7\pm0.3$\,$h^{-1}$\,Mpc,
which is similar to that derived in Geach et al.\ (2008) for a smaller sample.
Note that the effect of applying a different redshift distribution on $r_0$
corresponds to a scaling in amplitude of $\int dz_a(dn_a/dz_a)^2 / \int
dz_b(dn_b/dz_b)^2$.

Contamination by non-HAEs reduces the amplitude of the correlation function by
a factor $(1-f)^2$ where $f$ is the contamination fraction. As described in
\S2 it is likely that the contamination rate is of order 10\%, corresponding
to a factor 0.8 attenuation in the clustering amplitude. We do not apply a
correction to our measured parameters here until a more accurate estimate of
the contamination rate is obtained from our spectroscopic survey.

\begin{table*} \caption{Summary of model fit parameters to the observed
clustering of HAEs at $z=2.23$. Masses are in units of $h^{-1}M_\odot$ and
uncertainties reflect 1$\sigma$ range.}%\vspace{-0.5cm}

\begin{tabular}{cc}
  \hline
\multicolumn{2}{c}{Power-law$^a$} \cr
$r_0/({h^{-1}\,{\rm Mpc}})$ & $\chi^2/\nu$ \cr
\hline
$3.7\pm0.3$ & $0.9$ \cr
\hline
\end{tabular}

\begin{tabular}{ccc}
  \hline
\multicolumn{3}{c}{Dark matter$^{b}$} \cr
$\log_{10}(M_{\rm h})$ & $b_{\rm HAE}$ & $\chi^2/\nu$ \cr
\hline
 $11.7\pm0.1$ & $2.4^{+0.1}_{-0.2}$ & $0.8$ \cr
\hline
\end{tabular}

\begin{tabular}{ccccccc}
  \hline
 \multicolumn{7}{c}{Halo occupation distribution$^c$}\cr
 $\log_{10}(M_{\rm c})$ & $\log_{10}(M_{\rm eff})$ & $b_{\rm eff}$ & $f_{\rm sat}$ & $\sigma_{\log M}$ & $F_{\rm s}$ &$\chi^2/\nu$ \cr
\hline
$12.6^{+0.5}_{-1.6}$ & $12.1^{+0.1}_{-0.2}$ & $2.4^{+0.3}_{-0.4}$ & $0.08^{+0.37}_{-0.04}$ & $0.62^{+0.64}_{-0.60}$ & $0.3^{+0.7}_{-0.2}$ &  $0.7$\cr
\hline
\multicolumn{7}{l}{$^a$$\xi=(r/r_0)^{-1.8}$ fit for scales $\theta<600''$.}\cr
\multicolumn{7}{l}{$^b$$\xi_{\rm gal}=b_{\rm gal}^2\xi_{\rm DM}$. Mass is the `characteristic' halo mass for the quoted bias.}\cr
\multicolumn{7}{l}{$^c$See section 4.3 for further details. Note: $\sigma_{\log M}=\delta_{\log M}$, $\alpha=1$, $M_{\rm c}=M_{\rm min}$.}
\end{tabular}
\end{table*}

\subsection{Estimating the bias and characteristic halo mass of HAEs at
$\mathbf{z=2.23}$}

The autocorrelation function of galaxies can related to that of the underlying
dark matter via the linear bias: $\xi_{\rm DM}=b^2\xi_{\rm g}$. This arises
because galaxies forming in the peaks of a Gaussian random fluctuation field
will be clustered in a way that is biased to that of the dark matter. This
bias will depend on the details of galaxy formation relative to the underlying
matter density. It is therefore an important part of our understanding of a
particular galaxy population.

With an estimate for $\xi_{\rm DM}$, we can fit the observed projected angular
correlation function for the scaling $b^2$. To evaluate $\xi_{\rm DM}$ (or
rather, its projection, $\omega_{\rm DM}$), we follow the method described by
Hickox et al.\ (2012) and others (e.g.\ Myers et al.\ 2007; Coil et al.\ 2008)
that we briefly review here. First, the projected angular correlation function
of dark matter is derived by calculating the nonlinear dark matter power
spectrum, $\Delta^2_{\rm NL}(k,z)$, using the code {\sc halofit} (Smith et
al.\ 2003), assuming $\Gamma = \Omega_{\rm m}h = 0.21$ as the slope of the
initial fluctuation power spectrum. The projected correlation function
$\omega_{\rm DM}(\theta)$, averaged over the redshift distribution of the
HAEs, can then be calculated following Myers et al.\ (2007, equation A6),
which projects the power spectrum into the angular correlation function using
Limber's equation. The dark matter correlation function is shown in Figure\ 1.

We fit for the $b^2$ scaling that minimises a $\chi^2$ fit with the observed
HAE angular correlation function, yielding $b_{\rm HAE}=2.4\pm0.1$, with
reduced $\chi^2/\nu=0.8$, formally slightly poorer than the power law fit. The
characteristic halo mass $M$ is related to the bias through the
parameterisation $b = f(\nu)$ where $\nu$ is the ratio of the critical
threshold for spherical collapse to the r.m.s. density fluctuation for a mass
$M$: $\nu =\delta_c/\sigma(M)$. The function $f(\nu)$ for a given cosmology is
usually derived by fitting a form to the output of {\it N}-body simulations;
here we apply the function of Tinker et al.\ (2010) (assuming halos are all
200 times the mean density of the Universe). The Tinker et al. fitting
function is similar to that of Sheth, Mo \& Tormen (2001), but predicts
slightly larger $b$ for large $\nu$ and slightly lower $b$ for small $\nu$
(asymptoting to constant $b$ for low mass halos, and scaling as a power law
for high masses). For a bias of $b_{\rm HAE}=2.4^{+0.1}_{-0.2}$, we calculate
a characteristic halo mass of $\log_{10}(M_{\rm
h}/[h^{-1}M_\odot])=11.7\pm0.1$ at $z=2.23$. This characteristic $M_{\rm
halo}$ corresponds to the top-hat virial mass (see\ e.g.\ Peebles\ 1993 and
references therein), in the simplified case in which all objects in a given
sample reside in halos of the same mass. We note that this mass is
approximately equal to the `effective' halo mass derived from full HOD
modelling, as discussed in \S4.3, but differs from some prescriptions in the
literature which assume that sources occupy all halos above some minimum mass.
Given the halo mass function at $z\sim2$ (e.g.\ Tinker et al.\ 2008) the
derived minimum mass is typically a factor of $\sim$2 lower, for the same
clustering amplitude, than the characteristic mass quoted here.

\subsection{Comparison to models of galaxy formation}

{\sc galform} (Cole et al.\ 2000) is a successful semi-analytic model, or
rather a suite of models, that describe galaxy formation using simplified
prescriptions for the radiative cooling of gas within dark matter halos, star
formation and feedback (both through supernovae and active galactic nuclei
[AGN]), along with a hierarchical component for growth set by the merger
histories of the halos the galaxies occupy. The latter is achieved by coupling
semi-analytic models to large $N$-body simulations in which halos (usually
defined as regions within which the matter density is
$\Delta=200\times\bar{\rho}(z)$) can be identified and tracked (see Merson et
al.\ 2012 in preparation).

The main criticism levelled at semi-analytics is that they are
over-complicated with too many free (and uncertain) parameters. The counter
argument is that galaxy formation is inherently complex, and semi-analytics
serve as a tool for exploring the physics shaping the evolution of the galaxy
population below the resolution that can be achieved in numerical simulations;
these models can be refined as empirical results improve. Furthermore,
semi-analytic models are successful in reproducing many of the key features of
the galaxy population, including the shape and evolution of the luminosity
functions of stellar mass (see Baugh\ 2006 for a review).

We consider the clustering properties of HAEs within the Millennium Simulation
(Springel et al.\ 2005), generated from three different {\sc galform}
simulations: Bower et al.\ (2006; B06), Font et al.\ (2008; F08) and Lagos et
al.\ (2011, L11). The B06 model, which includes a recipe for AGN-driven
feedback in massive halos, successfully reproduces key features of the local
and distant galaxy population, including the black hole--bulge mass scaling at
$z=0$, the shape of the $b_{\rm J}$- and $K$-band luminosity functions at
$z=0$ (successfully reproducing the exponential turn down at high
luminosities) and the evolution of the stellar mass function of galaxies out
to $z\sim4.5$. Orsi et al.\ (2010) studied the clustering of HAEs in the B06
model to assess the relative merits of different selection techniques for the
construction of future galaxy redshift surveys. The F08 and L11 models are
based on B06, with the key improvements that: (a) F08 includes a more
realistic prescription for gas cooling within satellite galaxies which orbit
within massive halos, and (b) L11 implements a pressure-based star formation
law following Blitz \& Rosolowsky (2006), and a more refined model of the ISM.
We refer the reader to the respective articles that describe each model in
detail. The selection of HAEs in {\sc galform} is described by Orsi et al.\ (2010).

The predicted galaxy correlation functions are effectively identical in slope
and amplitude in all three models, with $r = 3.8$--$4.2$\,$h^{-1}$\,Mpc when
the amplitude of the real space correlation function is equal to unity
$\xi(r)=1$. The similarity between the predictions is perhaps not surprising,
given the similarities in the underlying galaxy formation models. This is in
reasonable agreement with the amplitude of the real space correlation function
estimated from the de-projection of the angular correlation function of real
HAEs. In Figure\ 2 we compare $\xi(r)$ measured directly from the simulations
to our power law and scaled dark matter models of the real HAE angular
correlation function. As Figure\ 2 shows, both power law and scaled dark
matter fits to the data almost exactly match the clustering strength of {\sc
galform} HAEs on scales $r>0.5$\,$h^{-1}$\,Mpc. {\sc galform} has less
clustering than scaled dark matter at smaller (single halo) scales. We explore
this in the next section, with a more sophisticated model of the clustering of
HAEs than simple using a scaled version of the dark matter correlation
function.

\begin{figure}\includegraphics[height=0.5\textwidth,angle=-90]{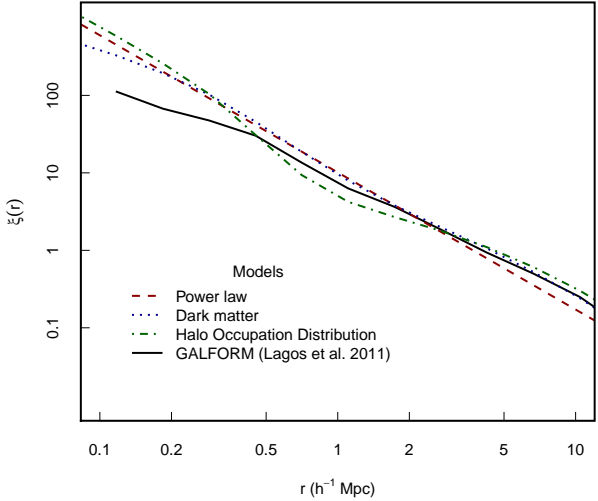}
\caption{A comparison of the real space correlation function of simulated HAEs
from {\sc galform} with $L_{\rm H\alpha}>10^{42}$\,erg\,s$^{-1}$\,cm$^{-2}$ at
$z=2.2$ to fits of the observed angular clustering (Fig\ 1). The lines show
three model fits to the measured angular correlation function: (a) a simple
power law $\xi(r)=(r/r_0)^{-\gamma}$ (with $\gamma=1.8$), (b) $\xi(r) =
b^2\xi_{\rm DM}$ and (c) the HOD fit (see \S4.3). On scales $\gsim$0.5\,Mpc
the models predict HAE clustering that is in reasonable agreement with the
amplitude of the clustering measured in the observations, but the
semi-analytic models predict less power at low separation compared to the data
(this is also apparent in the Bower et al.\ and Font et al. models which we do
not show here for clarity).} \end{figure}

\subsection{A Halo Occupation Distribution model for HAEs at
$\mathbf{z=2.23}$}

\subsubsection{Overview}

A basic tenet of our current picture of the formation of galaxies, and their
relationship to dark matter, is that galaxies inhabit dark matter halos either
as `central' galaxies close to the density peak, or `satellites' distributed
according to some radial density profile (Navarro, Frenk \& White\ 1997).
Intuitively, the number of satellites a halo can accommodate increases with
halo mass; illustrated in the real Universe by massive clusters of galaxies,
where the central galaxy is usually a massive elliptical surrounded by
hundreds or thousands of lower-mass cluster members. However, although the
occupation number might scale with halo mass in the stellar mass limited case,
the exact selection of galaxies in a given sample will affect the observed
halo occupation distribution. A halo occupation distribution (HOD) model
parameterises the probability distribution that describes the likelihood that
a halo of mass $M$ hosts on average $N$ galaxies (see Cooray \& Sheth\ 2002
for a review). As the projected clustering and number density of a galaxy
population (or populations) will depend on the form of the HOD, we can use the
observed clustering data to try to constrain models of the halo occupation of
HAEs. Critical to this approach is the parameterisation of the HOD; namely the
functional form assumed for the probability of finding a central galaxy, or
$N$ satellites in a halo of mass $M$.

We follow the methods of Wake et al.\ (2008, 2011 [W11]) to construct a halo
model, and refer the reader to Appendix B of W11 for a thorough description.
In brief, one must parameterise the halo model by defining functions for the
mean number of galaxies in a given halo, $\left<N|M\right>$. Given the good
agreement between the clustering amplitude measured from the semi-analytic
models and the data, we adjust our fiducial halo model to match the
simulations; here we have the luxury of the direct prediction of the HOD from
the model. In Figure\ 3 we show the HOD of $1.45\times10^7$ dark matter halos
in the Millennium Simulation, populated with HAEs using the {\sc galform}
model. We show the HAE HOD for three luminosity cuts, $L_{\rm
H\alpha}>10^{41}$, $10^{42}$, $10^{43}$\,erg\,s$^{-1}$.

The star-forming galaxy HOD has some important differences from typical mass
limited HODs (cf.\ Zheng et al.\ 2007, W11) that are worth noting. First, at
the lowest halo masses, the central galaxy distribution is approximately
Gaussian, with a characteristic host mass $M_{\rm min}$ and scale $\sigma$. At
halo masses $M\gsim M_{\rm min}+\sigma$ the distribution of centrals becomes
approximately flat, similar to the mass limited case though does not necessarily asymptote to $\left<N_{\rm c}|M\right>=1$. One could therefore
envisage a simple two component model for the central HAE halo occupation,
with a Gaussian distribution plus step function. At low H$\alpha$
luminosities, $L_{\rm H\alpha}\sim10^{41}$\,erg\,s$^{-1}$, above halo masses
of $\sim$$10^{11}h^{-1}\,M_\odot$ almost every halo hosts a central that is a
HAE. As the luminosity limit is increased, the low-mass Gaussian component
becomes more prominent (peaked) and shifted to higher halo masses, but with
the occupation number declining with increasing H$\alpha$ at all halo masses.

The decline in occupation number within increasing H$\alpha$ luminosity is in
part due to the form of the luminosity function, but the shape of the central
HOD is likely to be driven by (a) the stellar mass and star formation history
of central galaxies as a function of halo mass and (b) differences in the star
formation efficiency as a function of halo mass (e.g.\ the cooling rate onto
central galaxies). It is also important to consider that H$\alpha$ emission
can also result from nuclear activity which might be important for bright,
central HAEs in massive halos. The satellite distribution is similar to the
mass-limited case, with a smooth lower-mass cut-off in occupation and $\langle
N_{\rm s}|M\rangle$ scaling as a power-law at large $M$ (Kravtsov et al.\ 2004;
Zheng et al.\ 2005). There is a simple luminosity dependence, with the number
of satellites declining as $L_{\rm H\alpha}$ increases. The decline in
satellite occupation at all mass scales for the more luminous HAEs is a
natural outcome of the shape of the luminosity function, with $L_{\rm
H\alpha}=10^{43}$\,erg\,s$^{-1}$ probing exponentially declining $L>L^\star$
HAEs at this redshift (Geach et al.\ 2008; Sobral et al.\ 2012).

\begin{figure*} \includegraphics[width=0.33\textwidth,angle=-90]{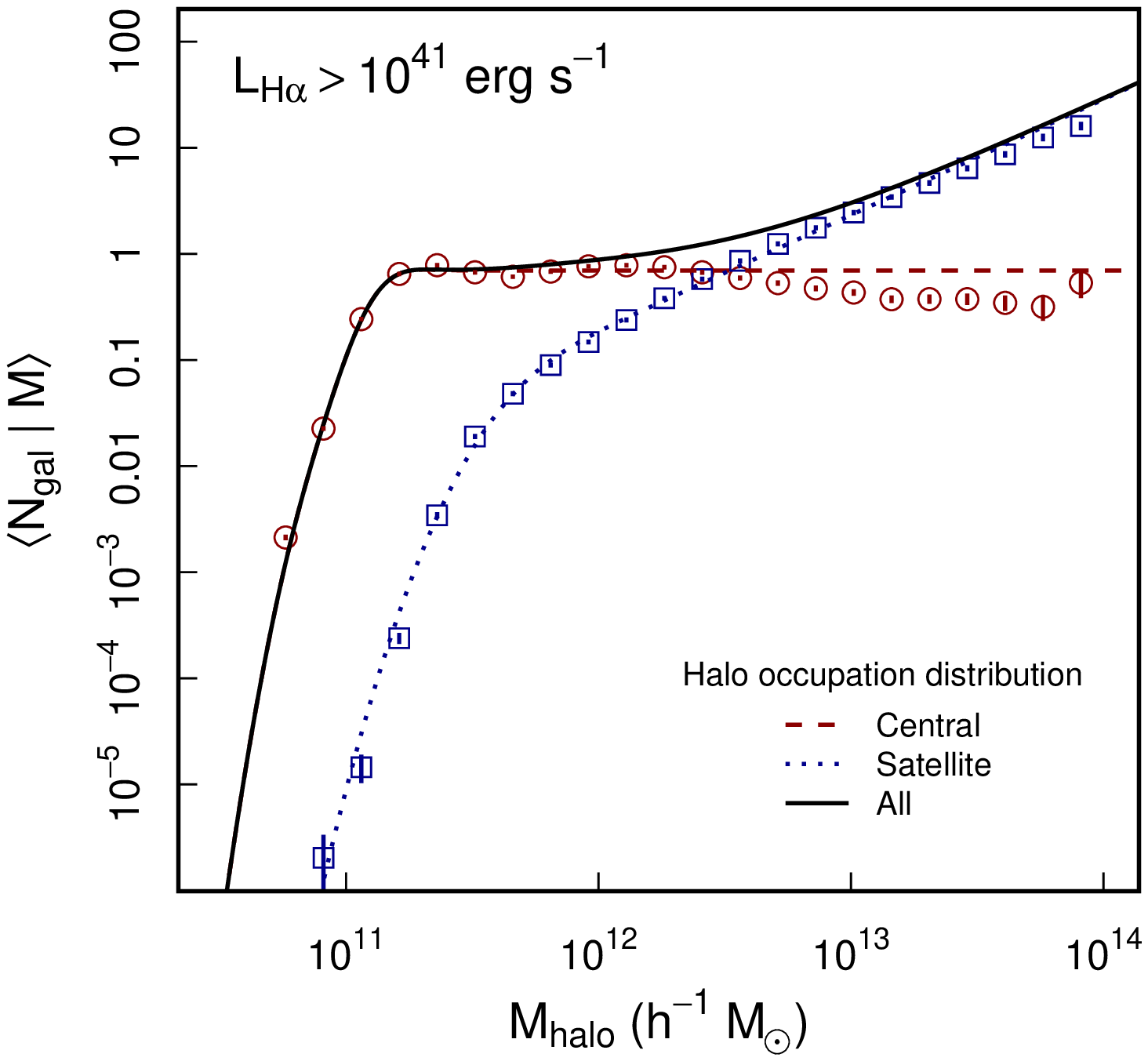}
\includegraphics[width=0.33\textwidth,angle=-90]{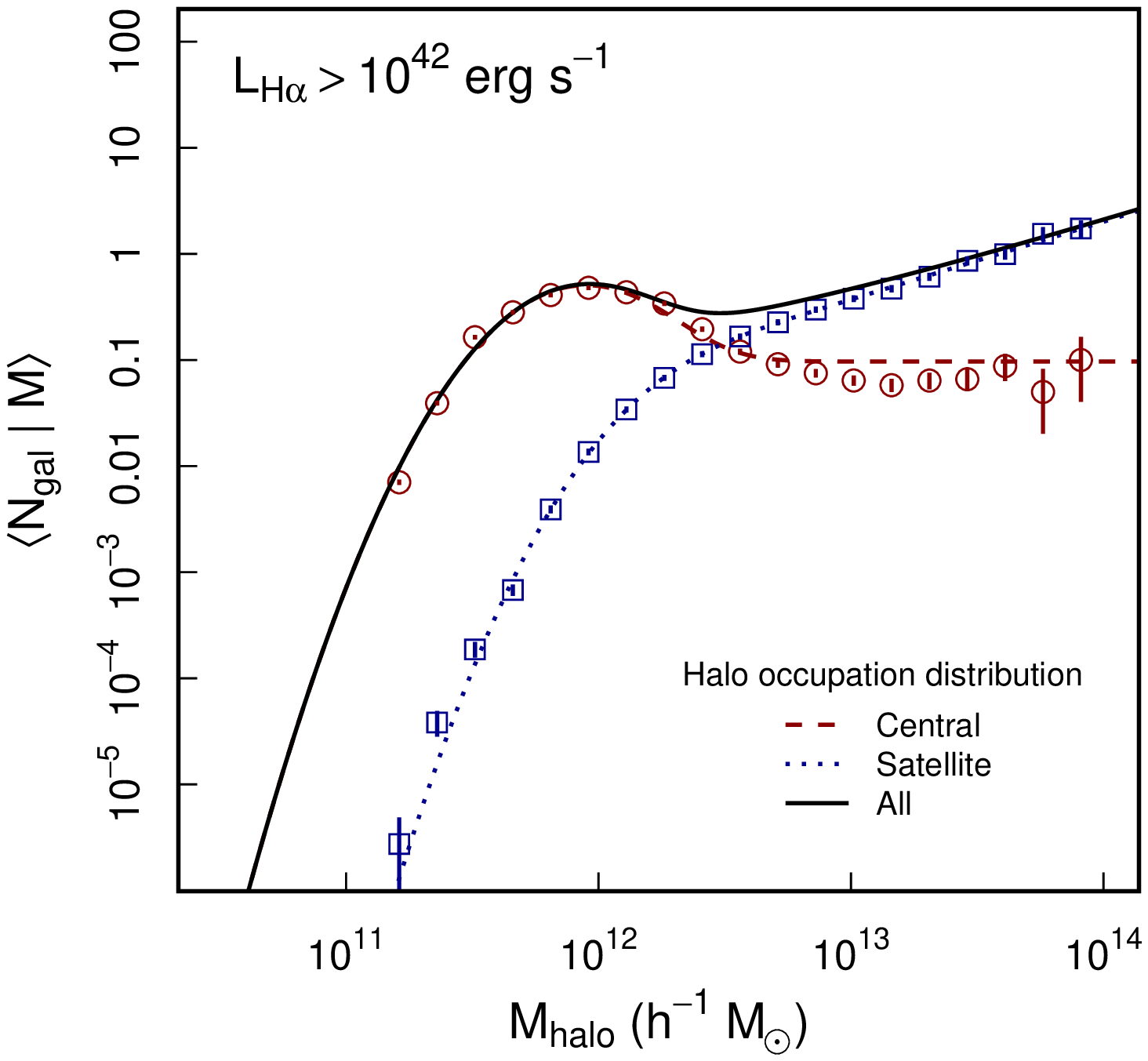}
\includegraphics[width=0.33\textwidth,angle=-90]{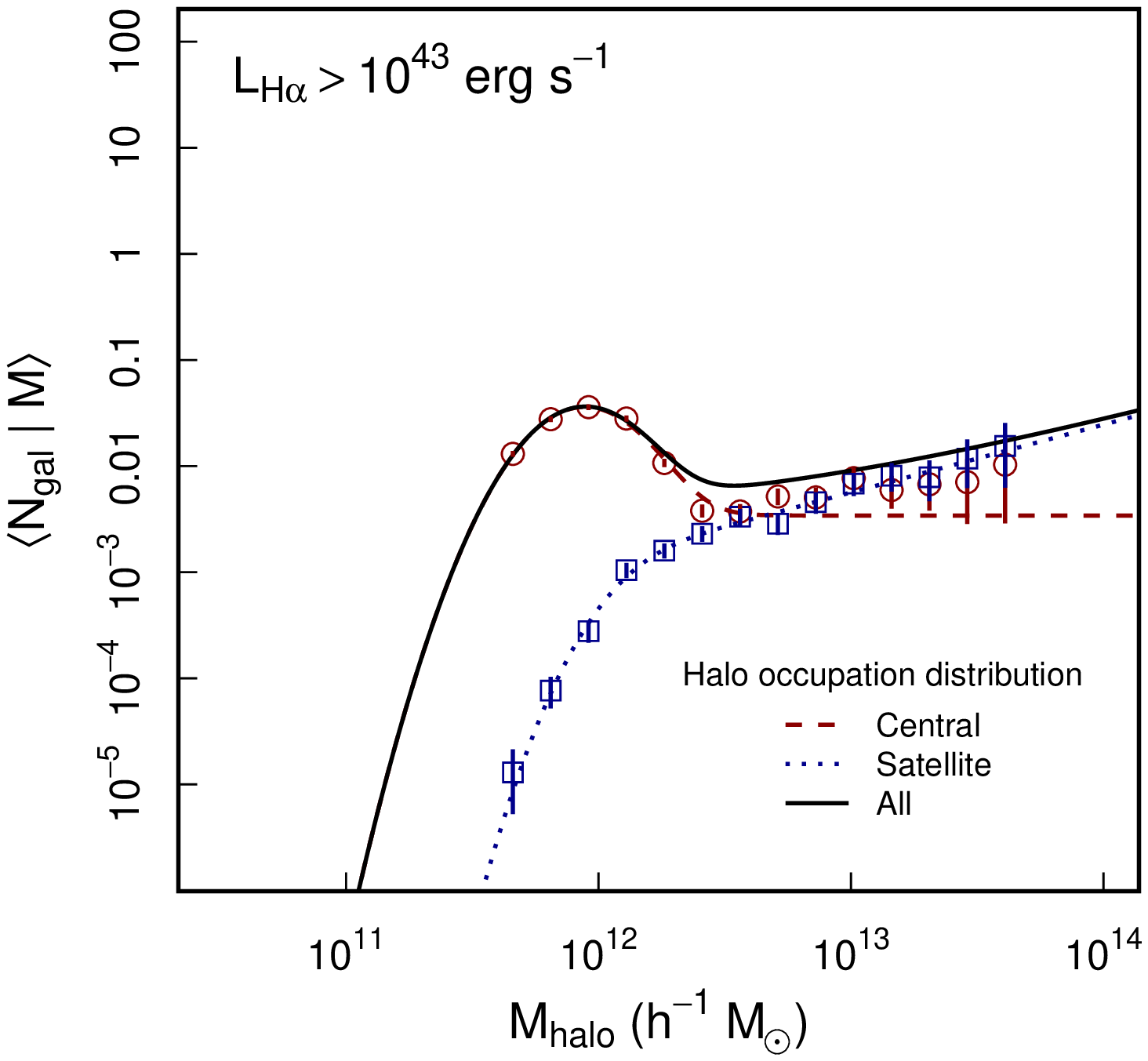} \caption{Halo
occupation distribution (HOD) model of HAEs at $z=2.2$ predicted by {\sc
galform}, where $\left<N_{\rm gal}|M\right>$ denotes the mean number of
galaxies in a halo of mass $M$. We show the HODs of central and satellite
galaxies with H$\alpha$ luminosities of (left to right panels) $L_{\rm
H\alpha}>10^{41},\,10^{42},\,10^{43}$\,erg\,s$^{-1}$ (points). The total
number of halos (that occupy the Millennium Simulation volume) in this model
is $1.45\times10^7$ (error bars are Poisson). There is a clear luminosity
dependence to the HOD, with the occupation number dropping at all halo masses
with increasing H$\alpha$ luminosity. The lines corresponding to `central',
`satellite' and `total' show the best fit to the points extracted from {\sc
galform} using our parametric HOD described in \S4.3. At all luminosities we
can fit the HOD with same parametric form, and we adopt this model in our
fitting of the observed projected correlation function.} \end{figure*}

\subsubsection{A HOD model for H$\alpha$ emitters}

The central HAE distribution can be adequately described by two components:

 \begin{align}\nonumber \left< N_{\rm c}|M\right> &=	 F^B_{\rm c}(1-F^A_{\rm c})\exp\left[ -\frac{\log
	(M/M_{\rm c})^2}{2\sigma^2_{\log M}}\right]\\
& +	F^A_{\rm c}\left[1+{\rm
	erf}\left(\frac{\log (M / M_{\rm c})}{\sigma_{\log M}}\right)\right]
\end{align}

\noindent where $F^{A,B}_{\rm c}$ are normalisation factors ranging from 0--1. The first
component describes the Gaussian distribution of centrals around halos of
average mass $M_{\rm c}$, and the second component describes the high mass
distribution, which we take as the standard mass-limited step function form
(Zheng et al.\ 2007). The parameter $\sigma_{\log M}$ describes the typical
mass range of halos with HAEs as centrals for the Gaussian component; the
exact value of $\sigma_{\log M}$ in the second component is not critical, and
so we decide to fix it to the Gaussian width. Similarly we set the step
function low mass cut-off to be $M_{\rm c}$. As shown in Figure\ 3, this four
parameter model provides a good description of the model HOD at
$10^{41}<(L_{\rm H\alpha}/{\rm erg\,s^{-1}})<10^{43}$, the pertinent range for
our analysis.

The number of satellite galaxies is described by a smoothed step function
similar to the central galaxy distribution for mass limited samples (Zheng et
al.\ 2007), but with the added component of a power law scaling at masses
larger than the critical mass, $M_{\rm min}$:

\begin{equation}
\left< N_{\rm s}|M\right> =
F_{\rm s}\left[1+{\rm erf}\left(\frac{\log (M / M_{\rm
min})}{\delta_{\log M}}\right)\right]\left(\frac{M}{M_{\rm min}} \right)^\alpha.
\end{equation}

\noindent The parameter $F_{\rm s}$ is the mean number of galaxies at the
transition mass $M_{\rm min}$ (the characteristic mass above which halos can
contain satellite HAEs). The parameter $\alpha$ controls the abundance of
star-forming satellites for $M>M_{\rm min}$. This functional form provides a
more satisfactory fit to the model satellite distribution at low masses
allowing a more gradual cut off to the power law than is assumed in the
standard stellar mass limited case (e.g.\ Wake et al.\ 2011). We make no
restrictions as to whether a central HAE is required for the hosting of
satellites, so the mean total number of galaxies in a halo of mass $M$ is

\begin{equation}
\left< N|M \right> = \left<N_{\rm c}|M\right> +\left< N_{\rm s}|M\right>.
\end{equation}

There are up to eight free parameters in this HOD. However we choose to fix
some in our modelling, given the size of the current sample. The exact
smoothing of the satellite low mass cut-off is not particularly important, in
that satellites close to the threshold (in the model) do not contribute
significantly to the halo occupation. Therefore we fix $\delta_{\log
M}=\sigma_{\log M}$. Although we do not require a halo to contain a H$\alpha$
emitting central in order to host satellite HAEs, we also constrain the
satellite threshold mass as $M_{\rm min}=M_{\rm c}$. Finally, we fix the slope
of the satellite distribution to $\alpha=1$; this is close to the model fit
across the full luminosity range shown in Figure\ 3, and is in agreement with
the value found for mass limited samples. Thus, our model has five free
parameters. Note that having a consistent model that scales with H$\alpha$
luminosity is of benefit to our analysis, given the possible uncertainties in
the fidelity of observed and simulated H$\alpha$ fluxes.

With $\left<N|M\right>$ defined, the number density of galaxies is given by
the integral of the halo mass function $n(M)$:

\begin{equation}
n_g = \int dMn(M)\left<N|M\right>,
\end{equation}

\noindent and this can be used as an additional constraint in the fitting of
the HOD, provided the number density of galaxies is known, although it is
often difficult to produce fits that simultaneously match the clustering and
abundance, e.g.\ Quadri et al.\ (2008). Here we use the latest
parameterisation for $n(M)$ from Tinker et al.\ (2010). With the halo model
set up, $\xi(r)$ is defined (Cooray \& Sheth\ 2002), and this can be projected
to the angular correlation function $\omega(\theta)$ using Limber's equation.

We can also define other parameters that are useful to summarize the halo
model: the satellite fraction,

\begin{equation}
f_{\rm sat} = \int dMn(M)\left<N_{\rm c}|M\right>\left<N_{\rm s}|M\right>/n_g,
\end{equation}

\noindent which measures the fraction of galaxies in the sample that are satellites; the effective halo mass:

\begin{equation}
M_{\rm eff} = \int dM M n(M)\left<N|M\right>/n_g,
\end{equation}

\noindent and the effective galaxy bias

\begin{equation}
b_{\rm g} = \int dMn(M)b(M)\left<N|M\right>/n_g,
\end{equation}

\noindent where $b(M)$ is the bias for a halo of mass $M$. 
 
\subsubsection{HOD fitting results}

We assert from the outset that, with the current data (i.e.\ relatively small
sample number), the interpretation of the results of this HOD analysis must be
taken with caution. Given the degeneracies involved, the results should only
be used as an early guide. Nevertheless, the HOD provides an elegant framework
within which to discuss the observed clustering, and we examine the results
here.

The angular correlation function derived from the HOD described above is fit
to the data, including the full covariance matrix. As in W11, minimisation is
achieved by using a Markov Chain Monte Carlo technique, which allows us to
efficiently explore the parameter volume. The best fit $\omega(\theta)$ is
shown in Figure\ 1, with a reduced $\chi^2/\nu=0.7$, again indicating that our
data is too coarse to constrain the model. Although we present the best
fitting model here, there are large degeneracies in the current halo model
that the data cannot resolve. This means that the key halo parameters
described in \S4.3.2 are only poorly constrained. The difference between the
HOD model and the real space correlation function measured from {\sc galform}
simulations is shown in Figure\ 2. Most of the parameters in equations 9 and
10 have very poor constraints, For example, the normalisation factors are
effectively unconstrained, and the 68\% confidence interval for the minimum
halo mass hosting centrals (and the minimum mass for satellites) is large,
$M_{\rm c}\sim$$(0.1$--$13)\times10^{12}h^{-1}\,M_\odot$ and the 1$\sigma$
upper limit of the satellite fraction $f_{\rm sat}\lsim0.46$. The
normalisation factors $F_c^{A,B}$ are effectively unconstrained.

There are clearly indications of serious degeneracies in the model that cannot
be resolved with the current data and are a common problem for samples of
galaxies where just a small fraction of the population are detected. Only the
average bias and mean halo mass are reasonably well constrained, with
$b=2.4_{-0.4}^{+0.3}$ and $M_{\rm
eff}=(1.3_{-0.5}^{+0.4})\times10^{12}h^{-1}\,M_\odot$, in agreement with what
was found for the scaled dark matter fit in \S4.1. We summarise the results
from the HOD fit in Table\ 1, along with the results from the power-law and
dark matter fits.

\begin{figure} \includegraphics[width=0.425\textwidth,angle=-90]{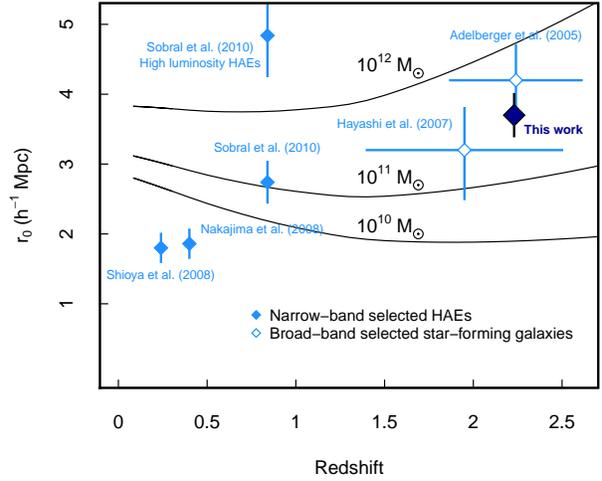}
\caption{Comparison of the correlation length of HAEs and star-forming
galaxies since $z=2.2$ derived from de-projected angular clustering
measurements. We compare the measured values to the predicted halo mass
hosting galaxies with correlation length $r_0$ for our cosmology. We
distinguish between measurements made from samples of HAEs selected in
narrow-band and more general star-forming galaxies selected in broad-band
surveys (the latter have much broader redshift distributions). Note that
evolutionary trends are hard to measure in this plot, given that the low
redshift surveys generally probe lower luminosity systems, and there is
observed to be a strong correlation between clustering strength and luminosity
(i.e.\ SFR). Indeed, Sobral et al.\ (2010) show that $r_0$ increases to
$r_0\sim5 h^{-1}$\,Mpc at $z=0.84$ when $L_{\rm
H\alpha}\gsim10^{42}$\,erg\,s$^{-1}$ are considered. In summary, `normal'
star-forming galaxies with SFR $\sim$1--100\,$M_\odot$\,yr$^{-1}$ have been
hosted by dark matter halos with $10^{10}\lsim M_{\rm
h}\lsim10^{12}\,h^{-1}M_\odot$ since $z=2.2$, with more luminous and massive
systems residing in more massive halos at all epochs.} \end{figure}

\section{Discussion}

\subsection{The fate of HAEs at $\mathbf{z=2.23}$}

The clustering amplitude of $z=2.23$ HAEs is similar to other star-forming
populations at high-{\it z}. Adelberger et al.\ (2005) present a clustering
analysis of $U_nG\mathcal{R}$ (BX/BM) selected star-forming galaxies at
$1.4<z<3.5$ and derive a correlation length of $r_0\sim4 h^{-1}$\,Mpc across
this redshift range, and argue that, at $z\sim2.2$, star-forming (BX) galaxies
with $M_\star\approx10^{10}M_\odot$ reside in dark matter halos of mass
$\sim$$10^{12}M_\odot$. Hayashi et al.\ (2007) present a clustering analysis
of star-forming `sBzK' selected galaxies (Daddi et al.\ 2004) at $z\sim2$,
which are a similar population to the broad-band BX selected galaxies
described above, finding $r_0 = 3.2^{+0.6}_{-0.7}$\,$h^{-1}$\,Mpc and typical
halo masses of $2.8\times10^{11}M_\odot$.

The average stellar mass of HAEs in our sample is $\log (M_\star/M_\odot)=9.4$
(calculated from stellar population fits to the homogenised UV-optical-near-IR
photometry using the templates of Bruzual \& Charlot\ 2007, including the
thermally pulsating Asymptotic Giant Branch population, Sobral et al.\ 2011).
The key improvement made here is that our selection is far more exclusive than
broad band selections, with the narrowband technique corresponding to a nearly
pure SFR selection over a very narrow redshift range. This has the effect of
minimising contamination (important for an accurate measurement of the
clustering amplitude for a specific population) and the tomographic nature of
the selection should improve the contrast of scale dependent features in the
projected clustering.

Hayashi et al. note the clear stellar mass ({\it K}-band luminosity)
dependence to the clustering strength, indicating that the descendants of sBzK
galaxies could range from sub-Milky Way mass halos to halos similar to rich
clusters. Sobral et al.\ (2010) also find that, when split by stellar mass and
H$\alpha$ luminosity, a clear increase in the derived correlation length was
found for HAEs at $z=0.84$, such that more massive and luminous (i.e.\ high
SFR) galaxies reside in more massive dark matter halos. The `varied fates' of
star-forming galaxies at $z=2$ has been discussed by Conroy et al.\ (2008) who
examine the evolutionary history of star-forming galaxy hosting dark matter
halos in {\it N}-body simulations, finding that generically selected
star-forming galaxies at $z\sim2$ do not evolve into any single class of
galaxy by $z=0$. The number density of the descendants of model $z\sim2$
star-forming galaxies at $z=0$ drops by a factor of two due to the merging of
descendants in the interval $0<z<2$. Of the remaining galaxies that did not
merge, 70\% evolve into central galaxies within halos of $M_{\rm
h}\gsim10^{12}h^{-1}M_\odot$ by $z=0$. Central galaxies at $z=0$ correspond to
$\gsim$$L^\star$ systems, whereas the star-forming galaxies that are destined
to become satellites by $z=0$ are generally lower-mass systems owing to the
slower/halted rate of stellar mass growth expected for sub-halos orbiting
within massive halos (i.e.\ a decline in the cooling rate and potential
expulsion of gas, q.v. \S4.2). Gonz\'alez et al.\ (2011) find a similar result
for submillimeter selected galaxies within {\sc galform}, with the descendants
of these high-{\it z} star-forming galaxies evolving into $z=0$ galaxies with
stellar masses $M_\star\sim10^{10-12}h^{-1}M_\odot$.

Although we expect the HAEs in our sample to evolve into a range of galaxy
types, we can estimate the halo mass of the descendants of the average HAE in
our sample -- i.e.\ those hosted by halos with the `characteristic' mass found
in our clustering analysis. Assuming $M_{\rm
eff}=(1.3_{-0.5}^{+0.4})\times10^{12}h^{-1}\,M_\odot$ at $z=2.23$ we use the
median halo mass growth rate from Fakhouri, Ma \& Boylan-Kolchin\ (2010) to
estimate that by $z=0$ the average HAE is destined to reside in a halo of mass
$M_{\rm h}=2$--$5\times10^{12}h^{-1}M_\odot$. Thus, HAEs are an important
population to study in the context of understanding the ecology of `typical'
galaxies in the local Universe, although as described above, there are likely
to be important mass and luminosity dependencies in the exact evolutionary
trajectory of HAEs (as hinted at by Figure\ 3 and 4), which our current data
cannot resolve.

\subsection{Comparison with other H$\alpha$ surveys at low redshift}

Sobral et al.\ (2010) present a clustering analysis of HAEs detected in HiZELS
at a redshift of $z=0.84$ (narrow {\it J}-band selection, probing to lower
H$\alpha$ luminosities than the present survey) and find a strong luminosity
dependence to the clustering strength, $2\lsim r_0 \lsim 5h^{-1}$\,Mpc for
$41.6<\log (L_{\rm H\alpha}/{\rm erg\,s^{-1}}) < 43.2$, with the clustering
strength increasing with luminosity (similar to the trend seen in other
samples, as described above). Our sample is too small to split into luminosity
bins and retain sufficient signal-to-noise in the clustering measurement. At
an equivalent luminosity limit to the one used in our analysis, the clustering
strength of HAEs at $z=0.84$ is similar to that at $z=2.23$, indicating only
weak evolution in the clustering properties of star-forming galaxies with
SFR$\gsim$10\,$M_\odot$\,yr$^{-1}$ over this range. Shioya et al.\ (2008) and
Nakajima et al.\ (2008) present clustering analyses for HAEs at $z=0.24$ and
$z=0.4$ respectively, finding correlation lengths of
$\sim$1.5--2\,$h^{-1}$\,Mpc. However, those studies probe fainter HAEs than
our sample contains, and therefore it is difficult to assess any redshift
evolution in the clustering properties of HAEs to these later epochs given the
expected strong luminosity dependence of $r_0$.

We summarise this comparison in Figure\ 4, where we compare the derived
correlation length of samples of narrow-band selected HAEs and the more
generic broad-band selections of star-forming galaxies described above. The
broad range in luminosity limits (Shioya et al.\ 2008 probe H$\alpha$
luminosities over two orders of magnitude lower than our sample) in the
$r_0$--$z$ plot mask any evidence of evolution in the clustering of
star-forming galaxies. Indeed, the characteristic luminosity of HAEs is itself
a strong function of redshift, with $\log (L^\star/{\rm erg\,s^{-1}}) = 0.45z
+ 41.87$ since $z=2.23$ (Sobral et al.\ 2012). It is clear however, that
`typical' star-forming galaxies (i.e.\ those close to $L^\star$ and not in the
ultraluminous class, such as submillimeter selected galaxies, see Hickox et
al.\ 2012) have, on average, been hosted by dark matter halos with
$10^{10}\lsim M_{\rm h}\lsim10^{12}\,h^{-1}M_\odot$ since $z=2.2$, with the
amplitude of clustering decreasing for less luminous and lower mass systems.

Figure\ 4 presents an average representation of the clustering properties of
star-forming galaxies. In reality, HAEs are expected to reside in halos with a
range of masses (as modelled by our HOD for example), and this will have
important consequences for their fate. In the next section we illustrate this
with an example from our data -- an apparent over-density of HAEs in the
COSMOS field, perhaps representing star-forming galaxies tracing a rather
massive, rare dark matter halo.

\begin{figure} \includegraphics[width=0.47\textwidth,angle=0]{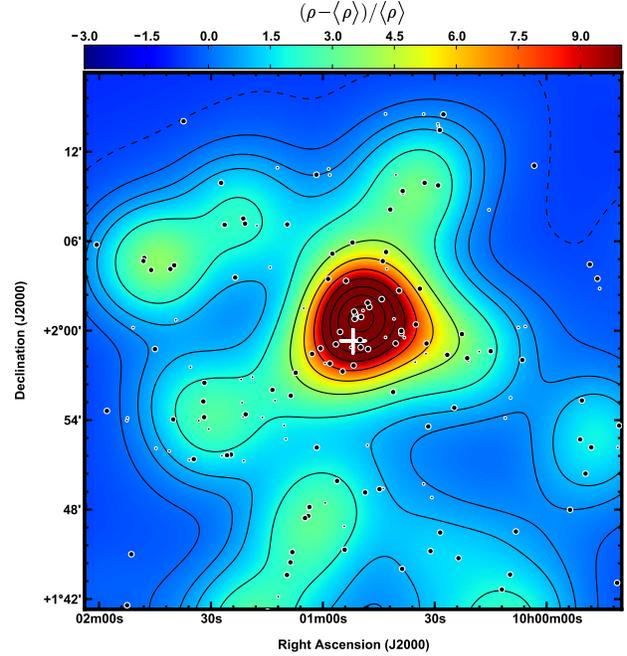}
\caption{A potentially massive halo in the COSMOS field, blindly detected as
an over-density of HAEs in HiZELS. The large points show HAEs meeting the
selection criteria used in the present study (smaller points are HAEs with
lower line fluxes). The colour background and contours show the smoothed
density contrast, $\delta = (\rho - \langle\rho\rangle)/\langle\rho\rangle$,
clearly indicating a significant peak in the mean surface density.
Interestingly, this structure contains a H$\alpha$ emitting $z=2.23$ QSO close
to the peak (cross symbol); such active systems are often used as `signpost'
objects around which to search for over-dense structures. HAEs in this
structure exemplify contribution of star-forming satellites producing power in
the correlation function at low angular separations.} \end{figure}

\subsection{A comment on satellite HAEs and cosmic variance: the detection of
a over-dense structure in the COSMOS field}

The measured correlation function implies that satellites play a
non-negligible role in the small-scale clustering power. In the halo model
described in \S4.3, massive halos with large numbers of bright
H$\alpha$--emitting satellites are rare objects, as dictated by the luminosity
and halo mass function. However such systems might be detectable in large
surveys such as ours as local over-densities in the surface density of HAEs.
We have detected such a system in the COSMOS field.

We have evaluated the local density contrast across the field by first
calculating a simple local density measure $\rho=4/\pi r^2_4$, where $r_4$ is
the angular distance to the fourth nearest HAE from an arbitrary point. This
is normalised to give the density contrast: $\delta = (\rho -
\langle\rho\rangle)/\langle\rho\rangle$, where $\langle\rho\rangle$ is the
mean surface density of HAEs across the field. We evaluate $\delta$ across a
grid, and then smooth this with a Gaussian kernel of {\sc fwhm} equivalent to
5 co-moving Mpc. The peak density contrast is $\delta=17$ at 10$^{\rm
h}$00$^{\rm m}$50$^{\rm s}$, +02$^\circ$00$'$53$''$. We do not detect a
similar structure in the UDS field, implying the sky density of environments
of similar mass is of the order one per two square degrees. Systems such as
this illustrate the importance of taking into account cosmic variance in
clustering measurements of HAEs. As Figure\ 1 shows, the small-scale
clustering power in the angular correlation function is dominated by the
COSMOS field, and this local over-density is likely to be a dominant
contributing factor, with the one-halo term boosting $\omega(\theta)$ at
scales below 1\,Mpc. The cosmic variance uncertainty is encoded into the
delete-one jackknife method we have employed, since the bulk of the
over-density is easily encompassed by one of the sub-volumes.

Figure\,5 shows the sky plot of HAEs around the peak of the over-density,
including a representation of the smoothed density field. Interestingly, the
peak encompasses the $z=2.2396$ quasar SDSS\,J100051.92+015919.2 (Prescott et
al.\ 2006), which is itself a HAE (and included in our sample). Extremely
luminous galaxies such as quasars and radio galaxies are often used to seek
out dense environments, relying on the fact that these extreme, but rare,
active galaxies are likely to be highly biased tracers of the matter field and
therefore reside in massive halos (Ellingson et al.\ 1991; Clowes \&
Campusano\ 1991; Bower \& Smail\ 1997; Miller et al.\ 2004; Boris et al.\
2007; Hatch et al.\ 2011; Matsuda et al.\ 2011). In this case the COSMOS
structure was blindly detected and turns out to harbour a quasar, lending
support for the approach of imaging the fields of active galaxies with
narrowband surveys to discover such (rare) environments.

\section{Summary}

We have presented an analysis of the clustering properties of 370 H$\alpha$
emitting galaxies at $z=2.23$, selected in two, independent, degree-scale
fields as part of the HiZELS survey. Using a series of increasingly
sophisticated models of the clustering, we find:

\begin{enumerate}

\item The average correlation function can be broadly modelled as a power law,
with slope $\beta=0.8$. Although there are clear deviations from the simple
power law on all scales, the normalisation of the power law fit provides an
adequate estimate of the physical correlation length of HAEs
$r_0=3.7\pm0.3$\,$h^{-1}$\,Mpc, similar to other star-forming populations at
this redshift. We find that the latest semi-analytic models of galaxy
formation predict a correlation length that is in good agreement with the
measured value.

\item The shape of the observed correlation function is more accurately
reproduced by scaling the projected correlation function of dark matter with a
bias factor: $\omega_{\rm HAE} = b^2\omega_{\rm DM}$. This is superior to the
simple power law as it is a better description of the variation of the power
in the correlation function across the full range of measured scales,
$0.1<r<10$\,$h^{-1}$\,Mpc. The best fitting value HAEs is $b_{\rm
HAE}=2.4^{+0.1}_{-0.2}$. This can be related to a characteristic halo mass,
which we find to be $\log (M_{\rm h}/[h^{-1}M_\odot]) = 11.7\pm0.1$.

\item Our final model attempts to fit the HAE clustering using a halo
occupation distribution (HOD) model. To parameterise the occupation of central
and satellite HAEs in dark matter halos, we turn to the semi-analytic models
for motivation (which predict the HOD), given the good agreement between model
described above. Although the HOD is poorly constrained by the current data
(with clear degeneracies resulting in multiple acceptable fits to the angular
clustering), we derive an average bias and characteristic halo mass in good
agreement with those derived from the scaled dark matter correlation function,
with $b=2.4_{-0.4}^{+0.3}$ and effective halo mass $M_{\rm
eff}=(1.3_{-0.5}^{+0.4})\times10^{12}h^{-1}\,M_\odot$.

\item Finally, we report on the detection of a significant localised
over-density of HAEs in the COSMOS field. Interestingly, this structure
encompasses a $z=2.23$ QSO, which is itself a HAE. It is clear from the
clustering analysis that cosmic variance in HAE surveys remains important on
$\sim$1\,deg$^2$ scales, especially in the fluctuations expected in the small
scale clustering amplitude. The HAE structure is likely to trace a relatively
massive halo, $M_{\rm h}\sim10^{13}M_\odot$ with a high satellite occupation
number, and could be destined to evolve into a large group or cluster of
galaxies by $z=0$. 

\end{enumerate}

Future high redshift H$\alpha$ surveys with improved statistics over wider
fields will be able to explore halo models of HAEs in further detail. Our
current result represents a first step in this direction, and despite the
limited information we can extract from the clustering models, it is clear
that disentangling the relative role of central and satellite star formation
in massive halos at high redshift is an important component of our
understanding of the efficiency of stellar mass assembly as a function of halo
mass. Multi-epoch H$\alpha$ surveys such as HiZELS will be essential for
examining evolutionary trends in the clustering properties of star-forming
galaxies at the peak era of galaxy formation, and we aim to investigate this
in a forthcoming paper.

\section*{Acknowledgements} JEG is supported by a Banting Fellowship,
administered by the Natural Science and Engineering Research Council (NSERC)
of Canada. DS acknowledges the award of a NOVA fellowship. IRS acknowledges
support from a Leverhulme Senior Fellowship and from the Science and
Technology Facilities Council (STFC). 

It is a pleasure to thank the entire team based at the Joint Astronomy Centre
who have made HiZELS a success. Particularly we would like to thank the
telescope operators Thor Wold, Tim Carroll and Jack Ehle and astronomers Luca
Rizzi, Tom Kerr and Andy Adamson. UKIRT will be sorely missed by the HiZELS
team.


\begin{thebibliography}{10}

\bibitem[Author et al. (2000)]{author}{Adelberger, K. L., Steidel, C. C.,
Pettini, M., Shapley, A. E., Reddy, N. A., Erb, D. K, 2005, ApJ, 619, 697}

 \bibitem[Author et al. (2000)]{author}{Baugh, C. M., 2006, A primer on
hierarchical galaxy formation: the semi-analytical approach, Reports on
Progress in Physics, 69, 3101--3156}

\bibitem[Author et al. (2000)]{author}{Benson, A. J., Cole, S., Frenk, C. S., Baugh, C. M., Lacey, C. G., 2000, MNRAS, 311, 793}

\bibitem[Author et al. (2000)]{author}{Best, P., Smail, I., Sobral, D., Geach,
J., Garn, T., Ivison, R., Kurk, J., Dalton, G., Cirasuolo, M., Casali, M.,
2010arXiv1003.5183}

\bibitem[Author et al. (2000)]{author}{Blakeslee et al. 2003, ApJ, 596, L143}

\bibitem[Author et al. (2000)]{author}{Blitz, L., Rosolowsky, E., 2006, ApJ, 650, 933}
	
\bibitem[Author et al. (2000)]{author}{Boris, N. V., Sodr\'e, L., Jr., Cypriano, E. S., Santos, W. A., de Oliveira, C. Mendes, West, M., 2007, ApJ, 666, 747}

\bibitem[Author et al. (2000)]{author}{Bower, R. G., Smail, Ian, 1997, MNRAS, 290, 292}

\bibitem[Author et al. (2000)]{author}{Bower, R. G., Benson, A. J., Malbon, R., Helly, J. C., Frenk, C. S., Baugh, C. M., Cole, S., Lacey, C. G., 2006, MNRAS, 370, 645}

\bibitem[Author et al. (2000)]{author}{Bruzual, G., 2007, From Stars to
Galaxies: Building the Pieces to Build Up the Universe. ASP Conference Series,
Vol. 374, Edited by Antonella Vallenari, Rosaria Tantalo, Laura Portinari, and
Alessia Moretti., p.303}

\bibitem[Author et al. (2000)]{author}{Clowes \& Campusano 1991, MNRAS, 249,
218 }

\bibitem[Author et al. (2000)]{author}{Cole, S., Lacey, C. G., Baugh, C. M., Frenk, C. S., 2000, MNRAS, 319, 168}

\bibitem[Author et al. (2000)]{author}{Colless, M, et al., 2001, MNRAS, 328, 1039}

\bibitem[Author et al. (2000)]{author}{Coil, A. L., et al.\ 2008, ApJ, 672,
153}

\bibitem[Author et al. (2000)]{author}{Conroy, C., Shapley, A. E., Tinker, J. L., Santos, M. R., Lemson, G., 2008, ApJ, 679, 1192}

\bibitem[Author et al. (2000)]{author}{Cooray, A., Sheth, R., 2002, PhR, 372, 1}

\bibitem[Author et al. (2000)]{author}{Daddi, E., Cimatti, A., Renzini, A., Fontana, A., Mignoli, M., Pozzetti, L., Tozzi, P., Zamorani, G., 2004, ApJ, 617, 746}

\bibitem[Author et al. (2000)]{author}{Elbaz, D., et al., 2011, A\&A, 533, 119}

\bibitem[Author et al. (2000)]{author}{Ellingson, E., Yee, H. K. C.; Green, R. F.,	1991, ApJS, 76, 455 }


\bibitem[Author et al. (2000)]{author}{Fakhouri O., Ma C.-P., Boylan-Kolchin M., 2010, MNRAS, 406, 2267}

\bibitem[Author et al. (2000)]{author}{Font, A. S., Bower, R. G., McCarthy, I. G., Benson, A. J., Frenk, C. S., Helly, J. C., Lacey, C. G., Baugh, C. M., Cole, S., 2008, MNRAS, 389, 1619}

\bibitem[Author et al. (2000)]{author}{Geach, J. E., Smail, Ian, Best, P. N., Kurk, J., Casali, M., Ivison, R. J., Coppin, K.,\ 2008, MNRAS, 388, 1473 }

\bibitem[Author et al. (2000)]{author}{Gonz\'alez, J. E., Lacey, C. G., Baugh, C. M., Frenk, C. S., 2011, MNRAS, 413, 749}

\bibitem[Author et al. (2000)]{author}{Groth, E. J., Peebles, P. J. E., 1977, ApJ, 217, 385}

\bibitem[Author et al. (2000)]{author}{Guaita, L., et al., 2010, ApJ, 714, 255}

\bibitem[Author et al. (2000)]{author}{Hall \& Green 1998, 507, 558 }

\bibitem[Author et al. (2000)]{author}{Hatch, N. A., et al.\ 2011, MNRAS, 410, 1537}


\bibitem[Author et al. (2000)]{author}{Hayashi, M., Shimasaku, K., Motohara, K., Yoshida, M., Okamura, S., Kashikawa, N., 2007, ApJ, 660, 72}

\bibitem[Author et al. (2000)]{author}{Hickox, R. C., et al.\ 2012, MNRAS, 421, 284}


\bibitem[Author et al. (2000)]{author}{Kennicutt, R., C., Jr., ARA\&A, 36, 189}

\bibitem[Author et al. (2000)]{author}{Kravtsov, A. V., Berlind, A. A., Wechsler, R. H., Klypin, A. A., Gottl\"ober, Stefan, Allgood, B., Primack, J. R., 2004, ApJ, 609, 35}

\bibitem[Author et al. (2000)]{author}{Lagos, C. Del P., Lacey, C. G., Baugh, C. M., Bower, R. G., Benson, A. J., 2011, MNRAS, 416, 1566}

\bibitem[Author et al. (2000)]{author}{Landy, S. D., \& Szalay, A.\ S., ApJ, 412, 64}
	
	\bibitem[Author et al. (2000)]{author}{Leureijs, R., 2011, arXiv:1110.3193}
	
\bibitem[Author et al. (2000)]{author}{Lawrence, A., et al., 2007, MNRAS, 379, 1599}

\bibitem[Author et al. (2000)]{author}{Limber, N.\ D., 1954, APJ, 119, 655}

\bibitem[Author et al. (2000)]{author}{Nakajima A., Shioya Y., Nagao T., Saito T., Murayam T., Sasaki S. S.,
Yokouchi A., Taniguchi Y., 2008, PASJ, 60, 1249}

\bibitem[Author et al. (2000)]{author}{Navarro, J. F., Frenk, C. S., White, S. D. M., 1997, ApJ, 490, 493}

\bibitem[Author et al. (2000)]{author}{Noeske, K. G, et al., 2007, ApJ, 660, 43}

\bibitem[Author et al. (2000)]{author}{Norberg, P., et al, 2001, MNRAS, 328, 64}

\bibitem[Author et al. (2000)]{author}{Norberg, P., Baugh, C. M., Gazta\~naga, E., Croton, D. J., 2009, MNRAS, 396, 19}

\bibitem[Author et al. (2000)]{author}{Matsuda, Y., et al., 2011, MNRAS, 416, 2041}

\bibitem[Author et al. (2000)]{author}{Mo, H. J., White, S. D. M., 1996, MNRAS, 282, 347}

\bibitem[Author et al. (2000)]{author}{Miller et al. 2004, MN, 355, 385 }

\bibitem[Author et al. (2000)]{author}{Myers, A.\ D., 2006, ApJ, 638, 622}

\bibitem[Author et al. (2000)]{author}{Myers, A. D., Brunner, R. J., Nichol, R. C., Richards, G. T., Schneider, D. P., Bahcall, N. A., 2007, ApJ 658, 85}


\bibitem[Author et al. (2000)]{author}{	Orsi, A., Baugh, C. M., Lacey, C. G., Cimatti, A., Wang, Y., Zamorani, G., 2010, MNRAS, 405, 1006}

\bibitem[Author et al. (2000)]{author}{Ouchi, M., et al., 2003, ApJ, 582, 60}

\bibitem[Author et al. (2000)]{author}{Peebles, P. J. E., 1980, The large-scale structure of the universe, Research supported by the National Science Foundation.~Princeton, N.J., Princeton University Press, 1980.~435 p.}

\bibitem[Author et al. (2000)]{author}{Peebles, P. J. E., 1993, Principles of physical cosmology, Princeton University Press, Princeton, NJ}

\bibitem[Author et al. (2000)]{author}{Prescott, M. K. M., Impey, C. D., Cool, R. J., Scoville, N. Z., 2006, ApJ, 644, 100}

\bibitem[Author et al. (2000)]{author}{Quadri, R. F., Williams, R. J., Lee, K.-S., Franx, M., van Dokkum, P., Brammer, G. B., 2008, ApJ, 685, 1}


\bibitem[Author et al. (2000)]{author}{Roche, N., Eales, S. A., Hippelein, H, Willott, C. J., 1999, MNRAS, 306, 538}
			
\bibitem[Author et al. (2000)]{author}{Ross, N. P., et al., ApJ, 697, 1634}
			
\bibitem[Author et al. (2000)]{author}{Rubin, V. C., 1954, Proceedings of the National Academy of Science, volume 40, 541--549}

\bibitem[Author et al. (2000)]{author}{Scoville, N., et al., 2007, ApJS, 172, 1}

\bibitem[Author et al. (2000)]{author}{Shao J., 1986, Ann. Stat., 14, 1322}

\bibitem[Author et al. (2000)]{author}{Sheth, Ravi K., Mo, H. J., Tormen, Giuseppe, 2001, MNRAS, 323, 1}

\bibitem[Author et al. (2000)]{author}{Shioya, Y., et al.\ ApJS, 175, 128}

\bibitem[Author et al. (2000)]{author}{Simon, P., 2007, A\&A, 473, 711}

\bibitem[Author et al. (2000)]{author}{Smith, R. E., Peacock, J. A., Jenkins, A., White, S. D. M., Frenk, C. S., Pearce, F. R., Thomas, P. A., Efstathiou, G., Couchman, H. M. P., 2003, MNRAS, 341, 1311}

\bibitem[Author et al. (2000)]{author}{Sobral, D., Smail, I., Best, P. N.,
Geach, J. E., Matsuda, Y., Stott, J. P., Cirasuolo, M., Kurk, J., 2012,
2012arXiv1202.3436S, MNRAS in press}

\bibitem[Author et al. (2000)]{author}{Sobral, D., Best, P. N.; Smail, Ian; Geach, J. E.; Cirasuolo, M.; Garn, T.; Dalton, G. B., 2011, MNRAS, 411, 675}

\bibitem[Author et al. (2000)]{author}{Sobral, D., Best, P. N., Geach, J. E.,
Smail, I., Cirasuolo, M., Garn, T., Dalton, G. B. Kurk, J., 2010, MNRAS, 404,
1551}

\bibitem[Author et al. (2000)]{author}{Spitler, L. R., et al., 2012, ApJL, 748, L21}

\bibitem[Author et al. (2000)]{author}{Springel, V., et al.\ 2005, Nature, 435, 629}

\bibitem[Author et al. (2000)]{author}{{Tinker}, J., {Kravtsov}, A.~V., {Klypin}, A., {Abazajian}, K., 
	{Warren}, M., {Yepes}, G., {Gottl{\"o}ber}, S., {Holz}, D.~E., 2008, ApJ, 688, 709}

\bibitem[Author et al. (2000)]{author}{Tinker, J. L., Robertson, B. E., Kravtsov, A. V., Klypin, A., Warren, A. S., Yepes, G., Gottl\"ober, S., 2010 ApJ, 724, 878}
	
\bibitem[Author et al. (2000)]{author}{Wake, D. A., et al., 2008, MNRAS, 387, 1045}

\bibitem[Author et al. (2000)]{author}{Wake, D. A., et al., 2011, ApJ, 728, 46}

\bibitem[Author et al. (2000)]{author}{White, S. D. M., Rees, M. J., 1978, MNRAS, 183, 341}

\bibitem[Author et al. (2000)]{author}{Yang, X., Mo, H. J., van den Bosch, F. C., 2003, MNRAS, 339, 1057}

\bibitem[Author et al. (2000)]{author}{York, D., et al.\ 2000, AJ, 120, 1579}

\bibitem[Author et al. (2000)]{author}{Zehavi, I., et al., 2011, ApJ, 736, 59}

\bibitem[Author et al. (2000)]{author}{Zheng, Z., Berlind, A. A., Weinberg, D. H., Benson, A. J., Baugh, C. M., Cole, S., Dav\'e, R., Frenk, C. S., Katz, N., Lacey, C. G., 2005, ApJ, 633, 791}
	
\bibitem[Author et al. (2000)]{author}{Zheng, Z., Coil, A. L., Zehavi, I., 2007, ApJ, 667, 760}

\end{thebibliography}
\end{document}